\documentclass{aa}

\usepackage{graphicx}
\usepackage{txfonts}
\usepackage[binary-units=true]{siunitx}
\usepackage{xspace}
\usepackage[pdftex]{hyperref}
\makeatletter
\renewcommand*\aa@pageof{, page \thepage{} of \pageref*{LastPage}}
\makeatother

\newcommand{\rsun}{R$_\odot$\xspace}
\newcommand{\lasco}{LASCO~C2\xspace}
\newcommand{\lya}{Lyman~$\alpha$\xspace}
\begin{document} 

   \title{Beyond the disk: EUV coronagraphic observations of the Extreme Ultraviolet Imager on board Solar Orbiter}
        \author{
        F. Auch\`ere\inst{\ref{i:ias}}
        \fnmsep\thanks{Corresponding author: Frédéric Auchère \email{frederic.auchere@universite-paris-scalay.fr}}
        \and
        D. Berghmans\inst{\ref{i:rob}}
        \and
        C. Dumesnil\inst{\ref{i:ias}}
        \and
        J.-P. Halain\inst{\ref{i:csl}}
        \and
        R. Mercier\inst{\ref{i:iogs}}
        \and
        P. Rochus\inst{\ref{i:csl}}
        \and
        F. Delmotte\inst{\ref{i:iogs}}
        \and
        S. Fran\c{c}ois\inst{\ref{i:ias}}
        \and
        A. Hermans\inst{\ref{i:csl}}
        \and
        V. Hervier\inst{\ref{i:ias}}
        \and
        E. Kraaikamp\inst{\ref{i:rob}}
        \and
        E. Meltchakov\inst{\ref{i:iogs}}
        \and
        G. Morinaud\inst{\ref{i:ias}}
        \and
        A. Philippon\inst{\ref{i:ias}}
        \and
        P. J. Smith\inst{\ref{i:mssl}}
        \and
        K. Stegen\inst{\ref{i:rob}}
        \and
        C. Verbeeck\inst{\ref{i:rob}}
        \and
        X. Zhang\inst{\ref{i:ias}}
        \and
        V. Andretta\inst{\ref{i:inaf-oac}}
        \and
        L. Abbo\inst{\ref{i:inaf-oato}}
        \and
        E. Buchlin\inst{\ref{i:ias}}
        \and
        F. Frassati\inst{\ref{i:inaf-oato}}
        \and
        S. Gissot\inst{\ref{i:rob}}
        \and
        M. Gyo \inst{\ref{i:pmod}}
        \and
        L. Harra\inst{\ref{i:pmod},\ref{i:eth}}
        \and
        G. Jerse\inst{\ref{i:inaf-ots}}
        \and
        F. Landini\inst{\ref{i:inaf-oato}}
        \and
        M. Mierla\inst{\ref{i:rob},\ref{i:geodin}}
        \and
        B. Nicula\inst{\ref{i:rob}}
        \and
        S. Parenti\inst{\ref{i:ias}}
        \and
        E. Renotte\inst{\ref{i:csl}}
        \and
        M. Romoli\inst{\ref{i:u-firenze}}
        \and
        G. Russano\inst{\ref{i:inaf-oac}}
        \and
        C. Sasso\inst{\ref{i:inaf-oac}}
        \and
        U. Sch\"uhle\inst{\ref{i:mps}}
        \and
        W. Schmutz\inst{\ref{i:pmod}}
        \and
        E. Soubrié\inst{\ref{i:ias}}
        \and
        R. Susino\inst{\ref{i:inaf-oato}}
        \and
        L. Teriaca\inst{\ref{i:mps}}
        \and
        M. West\inst{\ref{i:swri}}
        \and
        A.~N.~Zhukov\inst{\ref{i:rob},\ref{i:sinp}}
        }
        
        \institute{
            Université Paris-Saclay, CNRS, Institut d'Astrophysique Spatiale, 91405, Orsay, France\label{i:ias}
            \and
            Solar-Terrestrial Centre of Excellence -- SIDC, Royal Observatory of Belgium,  Brussels, Belgium\label{i:rob}
            \and
            Centre Spatial de Li\`ege, Universit\'e de Li\`ege, Av. du Pr\'e-Aily B29, 4031 Angleur, Belgium\label{i:csl}
            \and
            Laboratoire Charles Fabry, Institut d'Optique Graduate School, Universit\'e Paris-Saclay, 91127 Palaiseau Cedex, France\label{i:iogs}
            \and
            UCL-Mullard Space Science Laboratory, Holmbury St.\ Mary, Dorking, Surrey, RH5 6NT, UK\label{i:mssl}
            \and
            Physikalisch-Meteorologisches Observatorium Davos, World Radiation Center, 7260, Davos Dorf, Switzerland\label{i:pmod}
            \and
            ETH-Zurich, H\"onggerberg campus, HIT building, Z\"urich, Switzerland\label{i:eth}
            \and
            Max Planck Institute for Solar System Research, Justus-von-Liebig-Weg 3, 37077 G\"ottingen, Germany\label{i:mps}
            \and
            INAF -- Osservatorio Astronomico di Capodimonte, Napoli, Italy\label{i:inaf-oac}
            \and
            INAF - Osservatorio Astrofisico di Torino, Pino Torinese, Italy\label{i:inaf-oato}
            \and
            INAF - Osservatorio Astronomico di Trieste, Basovizza, Trieste, Italy\label{i:inaf-ots}
            \and
            Dipartimento di Fisica e Astronomia, Universit{\`a} di Firenze, Italy\label{i:u-firenze}
            \and
            Institute of Geodynamics of the Romanian Academy, Bucharest, Romania\label{i:geodin}
            \and
            Southwest Research Institute, 1050 Walnut Street, Suite 300, Boulder, CO 80302, USA\label{i:swri}
            \and
            Skobeltsyn Institute of Nuclear Physics, Moscow State University, 119992 Moscow, Russia\label{i:sinp}
}

   \date{Received ; accepted }

 
  \abstract
   {Most observations of the solar corona beyond 2~\rsun consist of broadband visible light imagery carried out with coronagraphs. The associated diagnostics mainly consist of kinematics and derivations of the electron number density. While the measurement of the properties of emission lines can provide crucial additional diagnostics of the coronal plasma (temperatures, velocities, abundances, etc.), these types of observations are comparatively rare. In visible wavelengths, observations at these heights are limited to total eclipses. In the ultraviolet (UV) to extreme UV (EUV) range, very few additional observations have been achieved since the pioneering results of the Ultraviolet Coronagraph Spectrometer (UVCS).}
   {One of the objectives of the Full Sun Imager (FSI) channel of the Extreme Ultraviolet Imager (EUI) on board the Solar Orbiter mission has been to provide very wide field-of-view EUV diagnostics of the morphology and dynamics of the solar atmosphere in temperature regimes that are typical of the lower transition region and of the corona.}
   {FSI carries out observations in two narrowbands of the EUV spectrum centered on \SI{17.4}{\nano\metre} and \SI{30.4}{\nano\metre} that are dominated, respectively, by lines of \ion{Fe}{IX/X}  (formed in the corona around \SI{1}{\mega\kelvin}) and by the resonance line of \ion{He}{II} (formed around \SI{80}{\kilo\kelvin} in the lower transition region). Unlike previous EUV imagers, FSI includes a moveable occulting disk that can be inserted in the optical path to reduce the amount of instrumental stray light to a minimum.}
   {FSI detects signals at \SI{17.4}{\nano\metre} up to the edge of its field of view (7~\rsun), which is about twice further than was previously possible. Operation at \SI{30.4}{\nano\metre} are for the moment compromised by an as-yet unidentified source of stray light. Comparisons with observations by the LASCO and Metis coronagraphs confirm the presence of morphological similarities and differences between the broadband visible light and EUV emissions, as documented on the basis of prior eclipse and space-based observations.}
   {The very-wide-field observations of FSI out to about 3 and 7~\rsun, without and with the occulting disk, respectively, are paving the way for future dedicated instruments.}

   \maketitle


\section{Introduction}

Remote sensing observations of the solar corona beyond 2~\rsun  mostly consist of visible light (VL) imagery from space coronagraphs. The Large Angle Spectroscopic Coronagraph~\citep[LASCO,][]{Brueckner1995} on board the Solar and Heliospheric Observatory~\citep[SOHO,][]{Domingo1995} has been providing several images per hour quasi-continuously since January 1996. The COR1 and COR2 instruments, part of the Sun Earth Connection Coronal and Heliospheric Investigation~\citep[SECCHI,][]{Howard2008} on board the Solar Terrestrial Relations Observatory (STEREO) have been in operation since January 2006. These observations are primarily used to study the macroscopic kinematics of the plasma and to derive the electron number density from Thomson scattering.

In addition to complementary measurements of the electron number density, measurements of the properties of emission lines from multiply ionized ions give access to essential quantities, such as the chemical composition of the plasma or the temperatures of the electrons and ions~\citep[see, e.g.,   ][and references therein]{Phillips2008, DelZanna2018}. However, observations of emission lines beyond 2~\rsun are relatively rare.

\begin{figure*}
    \centering
    \includegraphics[width=0.8\textwidth]{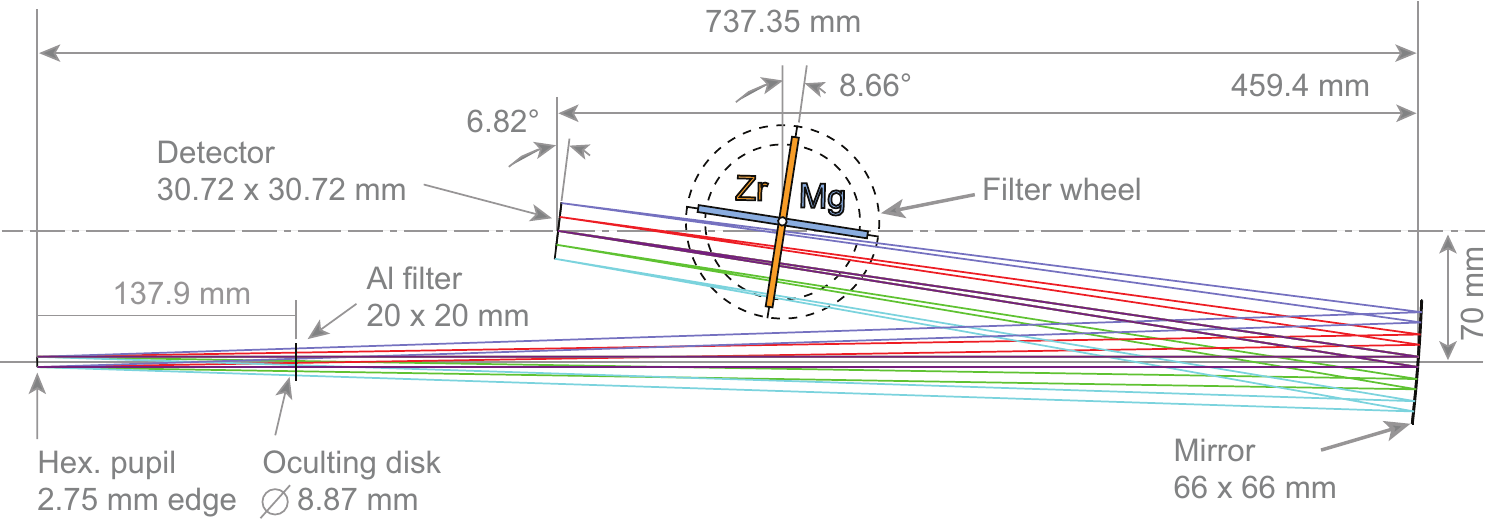}
    \caption{Optical layout of FSI~\citep[reproduced from ][]{Rochus2020}. The \SI{8.874}{\milli\metre} circular occulting disk is located on the entrance door (Fig.~\ref{fig:fsi_door}), \SI{135.9}{\milli\metre} behind the entrance aperture.}
    \label{fig:fsi_layout}
\end{figure*}

From the ground, visible and near-infrared coronal emission lines have been observed since the invention of the coronagraph~\citep{Lyot1932, Lyot1933}, however, due to the sky brightness, measurements beyond 2~\rsun are limited to total solar eclipses. During the 2006 March 29 eclipse, \cite{Habbal2007} observed the \ion{Fe}{XI} \SI{789.2}{\nano\metre} line out to 3~\rsun in streamers. \cite{Habbal2011} obtained simultaneous images with signal out to 2.4 to 3.4~\rsun in the \ion{Fe}{IX} \SI{435.9}{\nano\metre}, \ion{Fe}{X} \SI{637.4}{\nano\metre}, \ion{Fe}{XI} \SI{789.2}{\nano\metre}, \ion{Fe}{XIII} \SI{1074.7}{\nano\metre}, \ion{Fe}{XIV} \SI{530.3}{\nano\metre,} and \ion{Ni}{XV} \SI{670.2}{\nano\metre} spectral lines. Their formation temperatures, ranging from 0.5 to \SI{2.5}{\mega\kelvin,  } have allowed for detailed analysis and modeling \citep{Boe2022} of the temperature structure of the corona during the 2010 July 10 eclipse.

Without the limitation of the sky brightness, emission lines can be detected even further away from space. The Ultraviolet Coronagraph Spectrometer~\citep[UVCS,][]{Kohl1995a} on board SOHO provided groundbreaking spectroscopic UV observations from 1.2~\rsun to 10~\rsun in two channels centered on the \lya line of \ion{H}{I} and on the 103.2/\SI{103.7}{\nano\metre} doublet of \ion{O}{VI}. These were pioneered by an earlier version of UVCS during Spartan~201 flights~\citep{Kohl1994, Kohl1995b, Miralles1999}. We refer the reader to~\citet[][and references therein]{Kohl1997, Kohl2006} for reviews of results. The spectroheliographic soft X-ray imaging telescope (SPIRIT) on-board Coronas-F (in operation from 2001 to 2005) included a slitless grazing incidence spectrometer from \SIrange{28}{33.5}{\nano\metre} with a field of view (FOV) of 5~\rsun~\citep{Zhitnik2003a, Zhitnik2003b}. A similar instrument (TESIS) was flown on the Coronas-Photon mission~\citep{Kuzin2009, Kuzin2011}. These spectroheliographs were not coronagraphs, however, and were thus affected by instrumental stray light. Since the decommissioning of UVCS in 2012, no UV spectroscopic measurements have been made at these heights.

\begin{figure}
    \centering
    \includegraphics[width=\columnwidth]{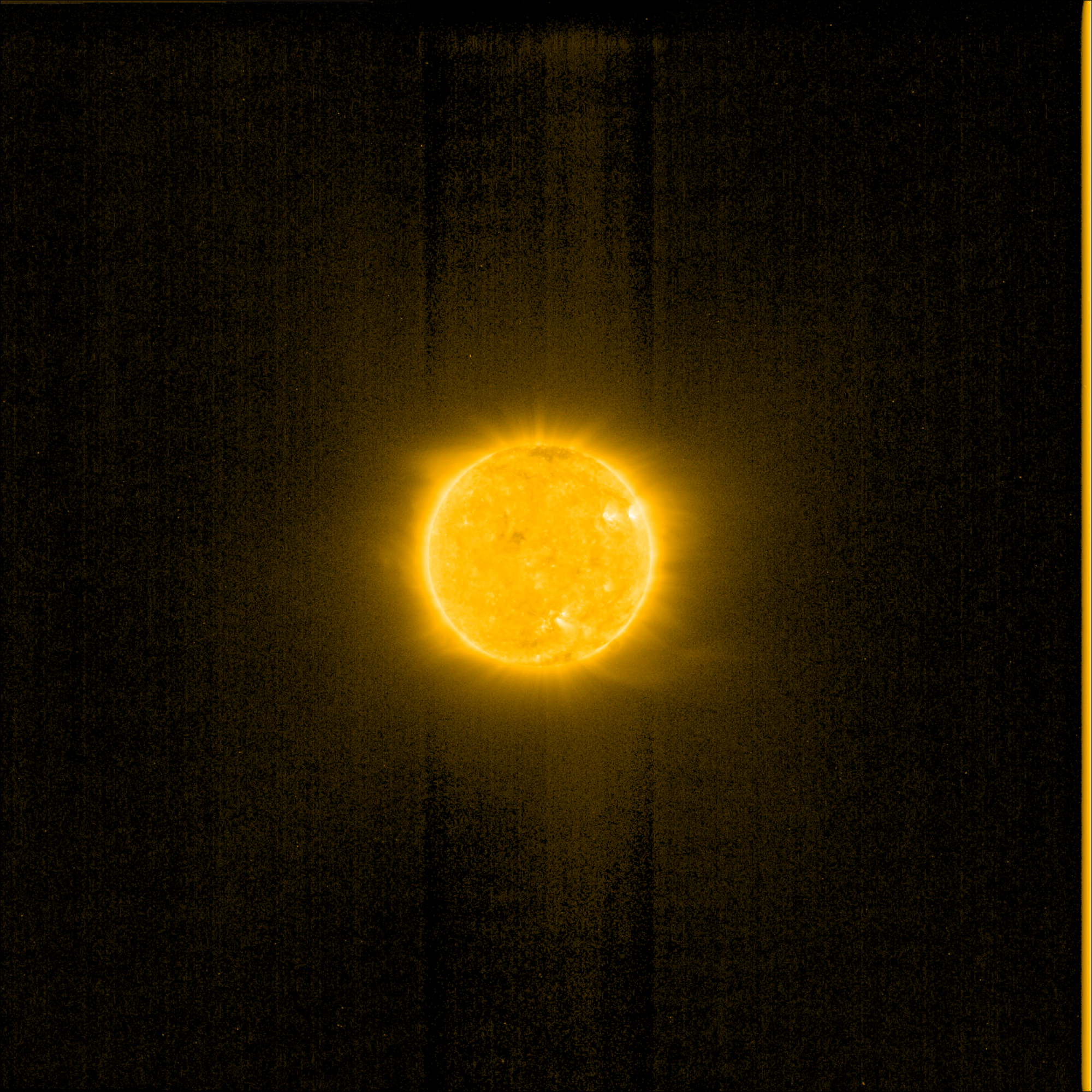}
    \caption{Regular \SI{10}{\second} exposure time \SI{17.4}{\nano\metre} FSI image taken at \SI{0.68}{\astronomicalunit} from the Sun on 2022 March 23 at 23:01. At this distance, the FOV extends to 4.9~\rsun. With this exposure time, signal is detected up to 2~\rsun. Longer exposures and the use of an occulting disk are required at larger angles (Fig.~\ref{fig:occulter_comparison}). The cause of the two vertical dark bands has not yet been identified.}
    \label{fig:fsi_disk}
\end{figure}

The narrowband imaging of emission lines from space beyond 2~\rsun is also very rare. During the first two years of the SOHO mission, the LASCO~C1 coronagraph had the capability to image up to 3~\rsun the \ion{Ca}{XV} (\SI{564.9}{\nano\metre}), \ion{Fe}{X} (\SI{637.4}{\nano\metre}), and \ion{Fe}{XIV} (\SI{530.3}{\nano\metre}) coronal lines~\citep{Schwenn1997}. SPIRIT included a narrowband normal incidence imager equipped with a moveable occulting disk and a steerable mirror to observe at 17.4 and \SI{30.4}{\nano\metre} up to 5~\rsun~\citep{Slemzin2008}. The HeCor EUV coronagraph~\citep{Auchere2007} made a single narrowband image at~\SI{30.4}{\nano\metre}  up to 3~\rsun during the first flight of the Herschel sounding rocket~\citep{Moses2020}.
Since 1996, the Extreme-ultraviolet Imaging Telescope \citep[EIT,][]{Delaboudiniere1995} on board SOHO and its successors provide regular narrowband EUV images, but the widest instantaneous FOV, that of the Extreme Ultraviolet Imager on board the STEREO~B spacecraft~\citep[EUVI,][]{Wuelser2007}, does not extend beyond 1.81~\rsun at \SI{1}{\astronomicalunit}. In addition, image compression frequently affects the data quality of EUVI in the outer field, which limits its practical FOV. SOHO has performed a dedicated offpoint maneuver on 1996 April 4, allowing EIT to observe up to 2.5~\rsun~\citep{Delaboudiniere1999, Slemzin2008}. The Sun Watcher with Active Pixels and Image Processing \citep[SWAP,][]{Seaton2013} and the Solar Ultra-Violet Imager~\citep[SUVI, ][]{Vasudevan2019} with half-FOVs of 1.69~\rsun and 1.67~\rsun respectively, have also performed offpoint maneuvers to explore the solar EUV corona up to about 3.5~\rsun at \SI{17.4}{\nano\metre}~\citep[SWAP, see ][]{West2022} and at 17.1 and \SI{19.3}{\nano\metre}~\citep[SUVI, see ][]{Takitonda2019, Seaton2021}. However, since EIT, SWAP, and SUVI are not coronagraphs, their observations at large angles from the limb are significantly affected by instrumental stray light.

The Solar Orbiter mission~\citep{Muller2020} includes two instruments capable of wide-field coronagraphic narrowband UV or EUV imaging: Metis~\citep{Antonucci2020} with its \lya channel, and the Extreme Ultraviolet Imager \citep[EUI,][]{Rochus2020}, with the Full Sun Imager (FSI) telescope. A description of the scientific objectives and associated observing programs of the payload can be found in~\citep{Zouganelis2020}. In Section~\ref{sec:full_sun_imager}, we describe FSI and, in particular, its capability to operate in coronagraphic mode. In Section~\ref{sec:campaigns}, We present the observations made so far by FSI in this particular mode during dedicated campaigns in chronological order. We compare the images qualitatively with simultaneous VL and UV coronagraphic observations made with Metis. Section~\ref{sec:conclusions} summarizes the results and presents a discussion of future developments.

\section{The Full Sun Imager: Need for an occulting disk\label{sec:full_sun_imager}}

\begin{figure*}
    \centering
    \includegraphics[width=\textwidth]{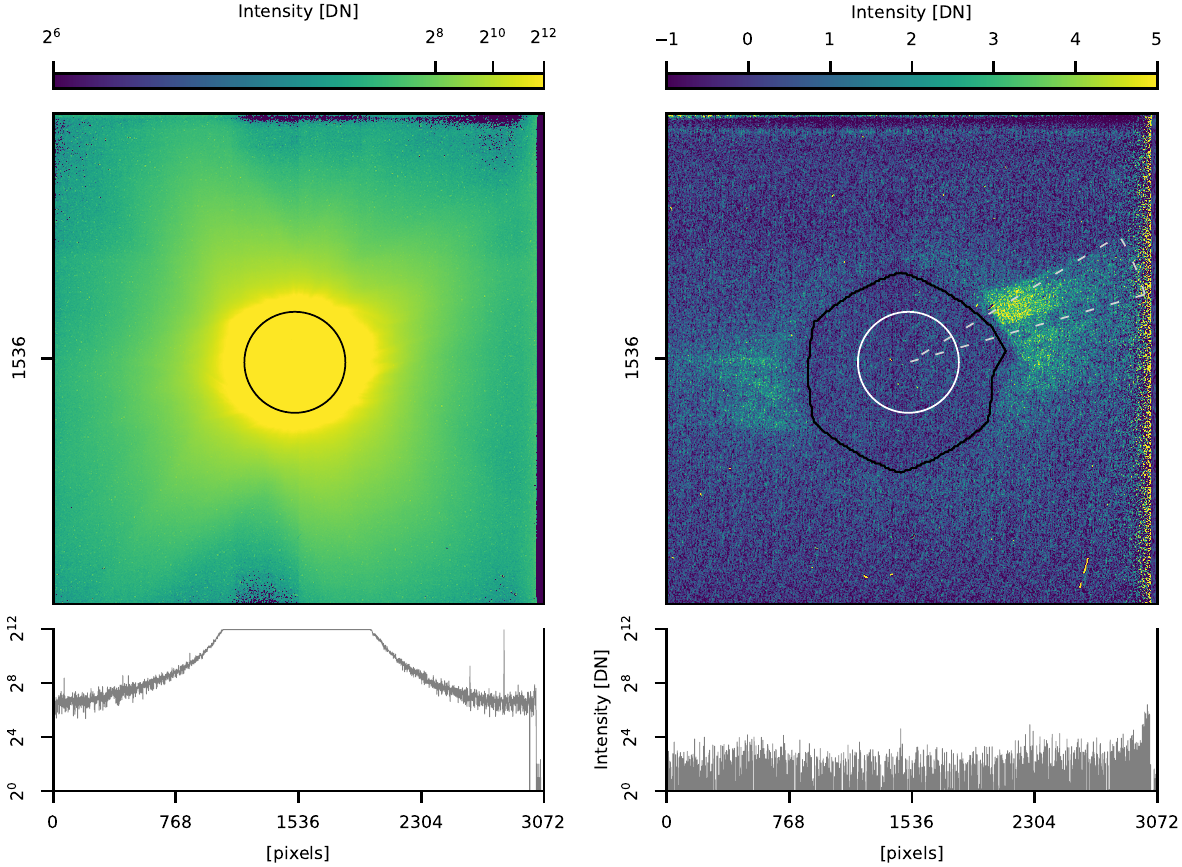}
    \caption{Effect of the occulting disk on \SI{640}{\second} FSI \SI{17.4}{\nano\metre} exposures taken on 2021 March 21. Left: View without the occulting disk at 01:48:45\,UT (logarithmic color scale). The position of the solar disk is marked by the white circle. Right: View with the occulting disk at 00:45:45\,UT (linear color scale). The black hexagonal shape marks the vignetting cut-off. A matching exposure dark frame was subtracted from each image. The intensity profiles at row 1536 are displayed on the same logarithmic scale for comparison.}
    \label{fig:occulter_comparison}
\end{figure*}

\begin{figure}
    \centering
    \includegraphics[width=\columnwidth]{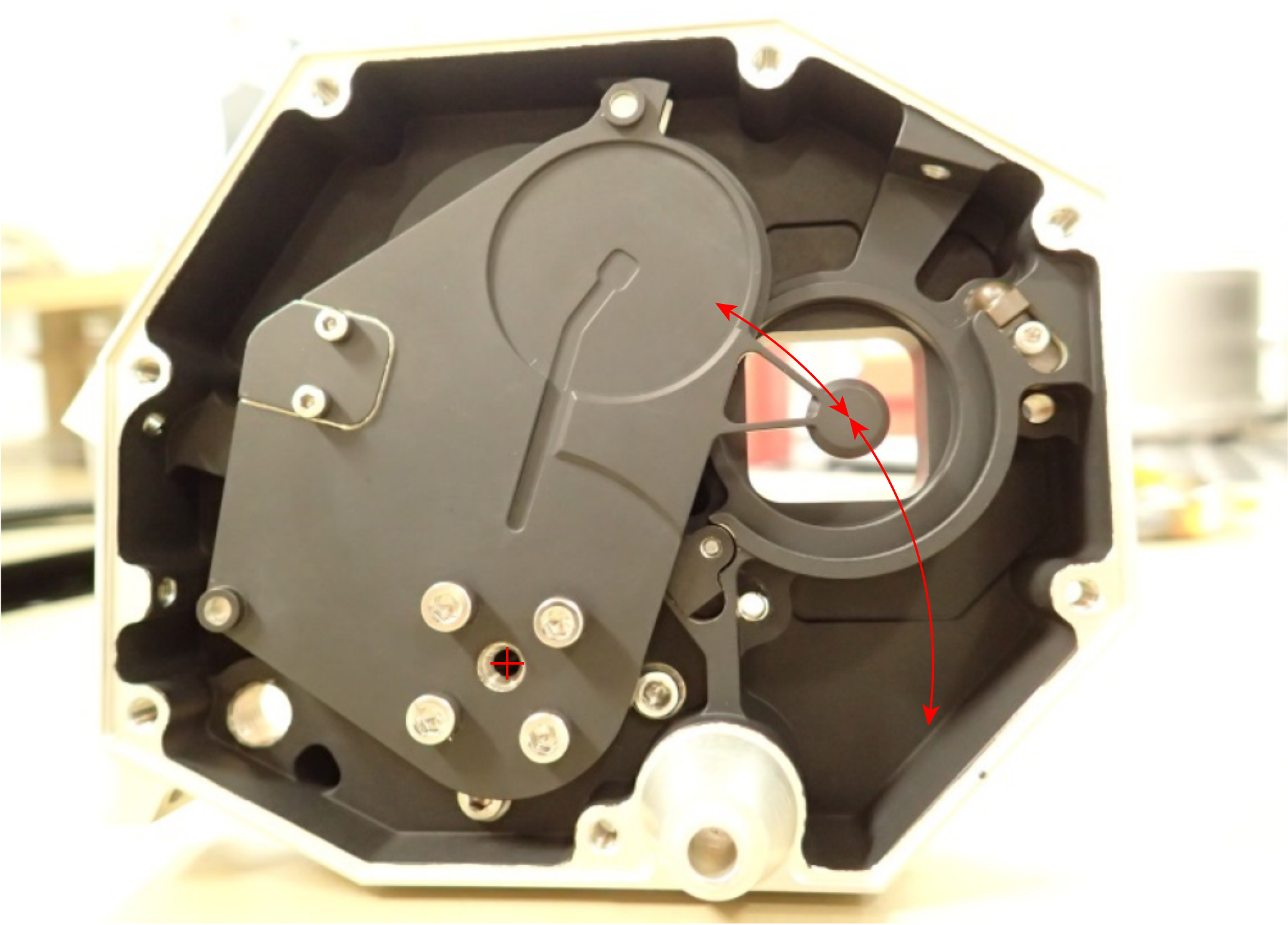}
    \caption{Entrance door of FSI, looking towards the Sun. The occulting disk is held off the door lid by two supporting rods. The occulting disk is shown in position. Rotating the lid around its axis (red cross) clockwise closes the door, rotating it counter-clockwise opens it. }
    \label{fig:fsi_door}
\end{figure}

The FSI is the wide field channel of the Extreme Ultraviolet Imager~\citep[EUI,][]{Rochus2020}, initially described in~\citet{auchere2005} and shown here in Figure~\ref{fig:fsi_layout}. It images the transition region and the corona in two narrowbands of the extreme ultraviolet (EUV) spectrum centered on 17.4 and \SI{30.4}{\nano\meter}, at an average plate-scale of \SI{4.46}{\arcsec} per \SI{10}{\micro\metre} pixel of the $3072\times3072$ active pixel sensor (APS). One of the purposes of FSI is  to serve as a context imager for the Solar Orbiter payload, thus, the two passbands were chosen to capture two major temperature regimes of the solar atmosphere. The \SI{17.4}{\nano\metre} passband (\SI{0.6}{\nano\metre} full width at half maximum) is centered on emission lines of \ion{Fe}{IX} and \ion{Fe}{X}, making it sensitive to coronal plasma around \SI{1}{\mega\kelvin}. This passband is very similar to the corresponding ones of EUVI and SWAP and is \SI{0.1}{\mega\kelvin} hotter than that of AIA \citep{Chen2021}. The \SI{30.4}{\nano\metre} passband (\SI{4}{\nano\metre} FWHM) is centered on the resonance line of \ion{He}{II}, formed around \SI{80}{\kilo\kelvin}, in the lower transition region. Over the mission duration, the orbital period of the spacecraft varies between 150 and 180 days, with the furthest aphelion at \SI{1.02}{\astronomicalunit} and the closest perihelion at \SI{0.28}{\astronomicalunit}. The FSI was designed with a $\SI{3.8}{\degree}\times$\SI{3.8}{\degree} FOV in order to cover two solar diameters at its closest approach, so that the whole disk can be seen even when the spacecraft is pointed at the limb. As a result, the half-FOV of FSI expressed in solar radii ranges from 2 to 7.25~\rsun depending on the distance to the Sun, much wider than that of any previous solar EUV imager (Fig.~\ref{fig:fsi_disk}). 

The typical exposure time of \SI{10}{\second} is set not to saturate the on-disk structures (Fig.~\ref{fig:fsi_disk}). In these regular exposures, except for prominence eruptions at \SI{30.4}{\nano\meter}~\citep{Mierla2022}, the signal beyond 2~\rsun is generally too faint to be detectable. Observations of the outer corona require longer exposure times. The left image of Fig.~\ref{fig:occulter_comparison} is a \SI{640}{\second} high-gain\footnote{EUI detectors are natively \SI{12}{\bit} and can be operated with two gains (high and low) that can either be used independently, or combined on board into \SI{15}{\bit} images. High-gain images have the lowest read noise.} exposure taken on 2021 March 21 at 01:48:45\,UT while Solar Orbiter was at \SI{0.68}{\astronomicalunit} from the Sun, giving a half-FOV of 4.9~\rsun. The image is completely saturated up to 1.2~\rsun. Solar structures are visible up to 2~\rsun but beyond, a diffuse haze of stray light dominates. It is caused by a mix of diffraction by the hexagonal mesh grid supporting the entrance aluminum filter (diagonal extensions) and of scattering by the mirror. The mesh grid used in FSI blocks the same fraction of the beam --  thus diffracting the same amount of light -- as those used in EUVI or in the Atmospheric Imaging Assembly~\citep[AIA,][]{Lemen2012}. The mirror substrate has a roughness of \SI{0.2}{\nano\metre}~rms and the multilayer coating has an interface roughness of \SI{0.5}{\nano\metre}~rms ~\citep{Meltchakov2013}, which are state-of-the-art values. Compared to Fig.~\ref{fig:fsi_disk}, the stray light in the left panel of Fig.~\ref{fig:occulter_comparison} is particularly prominent because the exposure is much deeper.

Measurements made in EIT images up to 1.8~\rsun during a transit of Mercury~\citep{Auchere2004} had already indicated that the solar signal in FSI (which uses the same technologies as EIT) would be dominated by instrumental stray light beyond a couple solar radii from the Sun's center. However, since very wide field imagery is not the primary objective of FSI, the addition of an occulting disk came late in the development of the instrument. It was introduced as a means of mitigating the descope of the \SI{30.4}{\nano\metre} channel originally present in Metis. The possibility to obtain \SI{30.4}{\nano\metre} wide field images with significant spatial overlap with the \lya (\SI{121.6}{\nano\metre}) channel of Metis enabled to preserve the scientific objective of making instantaneous maps of the abundance of Helium in the corona, using the method demonstrated by~\citet{Moses2020}.

\begin{figure}
    \centering
    \includegraphics[width=\columnwidth]{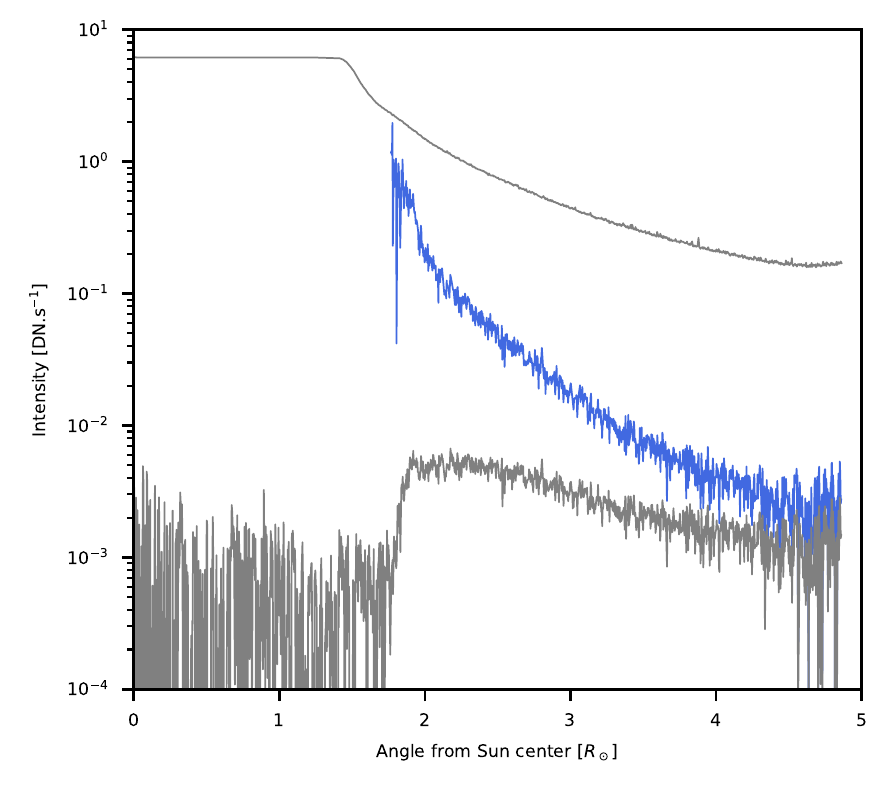}
    \caption{Raw (lower gray) and vignetting-calibrated (blue) radial \SI{17.4}{\nano\metre} intensity profiles, averaged over the sector shown in Fig.~\ref{fig:occulter_comparison}. The top curve corresponds to the data taken without the occulting disk.}
    \label{fig:vignetting_correction}
\end{figure}

Since the primary purpose of FSI is to observe the solar disk, the occulting disk has to be retractable. The decision was made, therefore, to implement it on the entrance door (as shown in Fig.~\ref{fig:fsi_door}). Since the door mechanism was originally not qualified for more than 200 operations and because of the necessity to see the solar disk most of the time, usage of the occulting disk was foreseen to be limited to dedicated campaigns. Once in flight, given the interest of the data obtained with the occulting disk, the decision was made to further qualify the door up to a total of 720 open-close cycles using an engineering model, which now allows for more frequent use. Tests made during the development of HeCor~\citep{Auchere2007}, which was a demonstrator for FSI, showed that a triple disk occulting system was overperforming. Instead, FSI uses a single occulting disk located after the entrance pupil, making it an inverted coronagraph, such as Metis. The occulting disk diameter was set for the vignetting cutoff to start at \SI{0.711}{\degree}.

The image in the right panel of Fig.~\ref{fig:occulter_comparison} was taken one hour before the one on the left, with the same exposure time and with the occulting disk in place. The brighter vertical band at the right edge of the image corresponds to a hotter area of the detector for which the dark correction is imperfect.  The black hexagonal shape indicates the computed vignetting cut-off which (at this date) was at 2~\rsun. It is not circular because the entrance pupil is hexagonal. The right-hand side bulge is caused by the two rods holding the disk. The stray light haze is completely gone. The remaining solar signal is at most 5~DNs, but streamers are now clearly visible beyond 2~\rsun. A radial profile averaged tangentially over the sector shown in the right panel of Fig.~\ref{fig:occulter_comparison} is shown in  Fig.~\ref{fig:vignetting_correction} (bottom curve). The blue curve is corrected from the vignetting function (computed by ray-tracing using as-built dimensions) and can thus be directly compared to the data taken without the occulting disk (top curve). The stray light-free intensity represents 10\% and 1\% of the non-occulted signal at 2~\rsun and 4.5~\rsun, respectively. The signal rise above 4.5~\rsun is caused by dark signal not perfectly corrected along the hotter right edge of the detector (right panel of Fig.~\ref{fig:occulter_comparison}), where the electrical strap is connected. 

The decrease of intensity with height can be used to investigate the formation process of the spectral lines that contribute to the passband. A number of studies have found evidence that resonant scattering plays a significant role in the formation of the \SI{17.1}{\nano\metre} resonance line of \ion{Fe}{X}~\citep{Schrijver2000, Slemzin2008, Goryaev2014}. However, the estimation of the amount of stray light is crucial in this respect and FSI is the first instrument to provide  images free of stray light at this wavelength. Photometric analysis of these images will be the subject of future research.

\section{Data processing\label{sec:data_processing}}

\subsection{FSI}

The FSI images shown in this paper are based on the Level-1 data files published as part of the EUI Data Release 5.0 \citep{euidatarelease5}, except for the 2022 December data. Unless stated otherwise, all the coronagraph mode FSI images presented in this paper have been processed in the same way. First, a master dark frame with matching exposure time was subtracted. For the \SI{1000}{\second} exposures, the master dark frame was created from eight dark frames taken on 2021 November 2 and 4. In order to remove local spikes (e.g., cosmic rays) in the dark frame, we used ten iterations of 2~$\sigma$-clipping, followed by four iterations of a spatial $13\times 13$ 5~$\sigma$-clipping. For the \SI{640}{\second} exposures, a single dark frame was available, so only a spatial filtering  was applied. Spikes in the dark-subtracted images were removed using ten iterations of  seven temporal points running a median-centered 2~$\sigma$-clipping, followed by four iterations of a spatial $13\times 13$ 5~$\sigma$-clipping. For isolated exposures (e.g., the one shown in Fig.~\ref{fig:occulter_comparison}), only spatial filtering was applied. Variations in the detector offset between acquisitions causes vertical banding in the dark-subtracted frames. The bands are estimated in each image by averaging detector rows 50 to 250 and 2700 to 2900 followed by linear interpolation in the vertical dimension. Finally, the data cubes were denoised using the wavelet-based method described in \citet{Starck1994, Murtagh1995, Auchere2023}.

\subsection{Metis\label{sec:metis_data}}

Each Metis VL polarized frame was acquired with a $2\times2$ binning and an exposure time of \SI{30}{\second}. A set of 14 frames at the same polarization angle was averaged on board. During the acquisition, the polarizer was cycling through four polarization angles to create a quadruplet of polarized images, which were finally combined on ground to obtain a single polarized brightness (pB) image with an effective exposure time of \SI{1680}{\second}. During the November 2021 campaign, Metis was also operating its UV \lya channel. For those dates, the UV frames were acquired cotemporally with the pB sequence, with a $4\times4$ binning and exposure time of \SI{60}{\second}, at a cadence of \SI{120}{\second}. The UV data cube was $\sigma$-clipped with the same parameters as FSI to remove cosmic spikes and the images were averaged together by groups of ten to increase the signal-to-noise ratio.

The VL and UV images were processed and calibrated on the ground as described in \cite{Romoli2021}, with an updated radiometric calibration described in \cite{Andretta2021} and in \cite{DeLeo2023a, DeLeo2023b}. A dark and bias frame was subtracted from the images, which were then corrected for flat-field and vignetting, and normalized by the exposure time. A radiometric calibration was also applied, which takes into account a revised in-flight calibration obtained from a set of VL and UV standard stars \citep{DeLeo2023a, DeLeo2023b}. The bias and dark images for both detectors were acquired in flight, whereas the flat-field and vignetting images were measured on ground during the laboratory calibrations \citep{Antonucci2020}. In the particular case of UV images, the individual frames have been then corrected using the dark frames acquired closest in time with matching acquisition parameters. However, a residual variation in the dark level on short time scales is still noticeable in these data sets. We therefore applied a further correction to the UV dark levels, which is described in \cite{DeLeo2023b} and \cite{Russano2023}.

\begin{table*}
\center
\caption{Summary of the FSI coronagraph mode campaigns.}
\begin{tabular}{l@{ }ll@{ }lcccccc}
\hline\hline
 & & & & & & & & \multicolumn{2}{c}{Separation angle}\\ \cline{7-8}
\multicolumn{2}{l}{Start date (UT)} & \multicolumn{2}{l}{End date (UT)} & Channel & Exposure & Cadence & Sun distance & Earth & STEREO A\\
\hline
2021 Sep. 9 & 00:42 & 2021 Sep. 9 & 09:30 & \SI{17.4}{\nano\metre} & \SI{640}{\second} & \SI{11}{\minute} & \SI{0.60}{\astronomicalunit} & \SI{65}{\degree} & \SI{24}{\degree}\\
2021 Nov. 1 & 00:42 & 2021 Nov. 3 & 23:42 & \SI{17.4}{\nano\metre} & \SI{1000}{\second} & \SI{30}{\minute} & \SI{0.83}{\astronomicalunit} & \SI{2}{\degree} & \SI{36}{\degree}\\
2021 Nov. 4 & 00:12 & 2021 Nov. 4 & 21:12 & \SI{30.4}{\nano\metre} & \SI{1000}{\second} & \SI{30}{\minute} & \SI{0.84}{\astronomicalunit} & \SI{2}{\degree} & \SI{36}{\degree}\\
2022 Feb. 8 & 04:15 & 2022 Feb. 8 & 07:45 & \SI{17.4}{\nano\metre} & \SI{1000}{\second} & \SI{30}{\minute} & \SI{0.79}{\astronomicalunit} & \SI{19}{\degree} & \SI{16}{\degree}\\
2022 Mar. 7 & 16:00 & 2022 Mar. 7 & 19:30 & \SI{17.4}{\nano\metre} & \SI{1000}{\second} & \SI{30}{\minute} & \SI{0.50}{\astronomicalunit} & \SI{3}{\degree} & \SI{33}{\degree}\\
2022 Dec. 5 & 04:00 & 2023 Jan. 1 & 22:15 & \SI{17.4}{\nano\metre} & \SI{1000}{\second} & \SI{30}{\minute} & \SIrange{0.83}{0.95}{\astronomicalunit} & \SIrange{16}{22}{\degree} & \SIrange{4}{9}{\degree}\\
\hline
\end{tabular}
\label{tab:campaigns}
\end{table*}

\section{Campaigns\label{sec:campaigns}}

The door mechanism holding the occulting disk suffered from a positioning repeatability issue during the first year of the mission. Only a few test images were taken prior to that shown in the right panel of Fig.~\ref{fig:occulter_comparison}, which is the first one obtained after the solution was found.
Since then, the instrument has been run in coronagraph mode during dedicated campaigns, either in support of specific observations (e.g. Herschel sounding rocket flight, Sect.~\ref{sec:campaign032022}), or at times of upper conjunction with either Earth or the STEREO A spacecraft (Sect.~\ref{sec:campaign112021} and~\ref{sec:campaign122022}). Indeed, in this latter case, the FSI images in disk mode provide mostly redundant information, as compared to those from other EUV imagers in operation (EIT, AIA, SWAP, SUVI, and EUVI). Table~\ref{tab:campaigns} lists the main characteristics of each campaign. All \SI{30.4}{\nano\metre} images obtained so far in coronagraph mode exhibit an unexplained parasitic pattern, the intensity of which is comparable to that of the expected signal, so they are not used in this work. Except for test images during the November 2021 campaign and until the source of the parasite is identified and suppressed, all coronagraph mode images were taken in the \SI{17.4}{\nano\metre} passband. 

\begin{figure}
    \centering
    \includegraphics[width=\columnwidth]{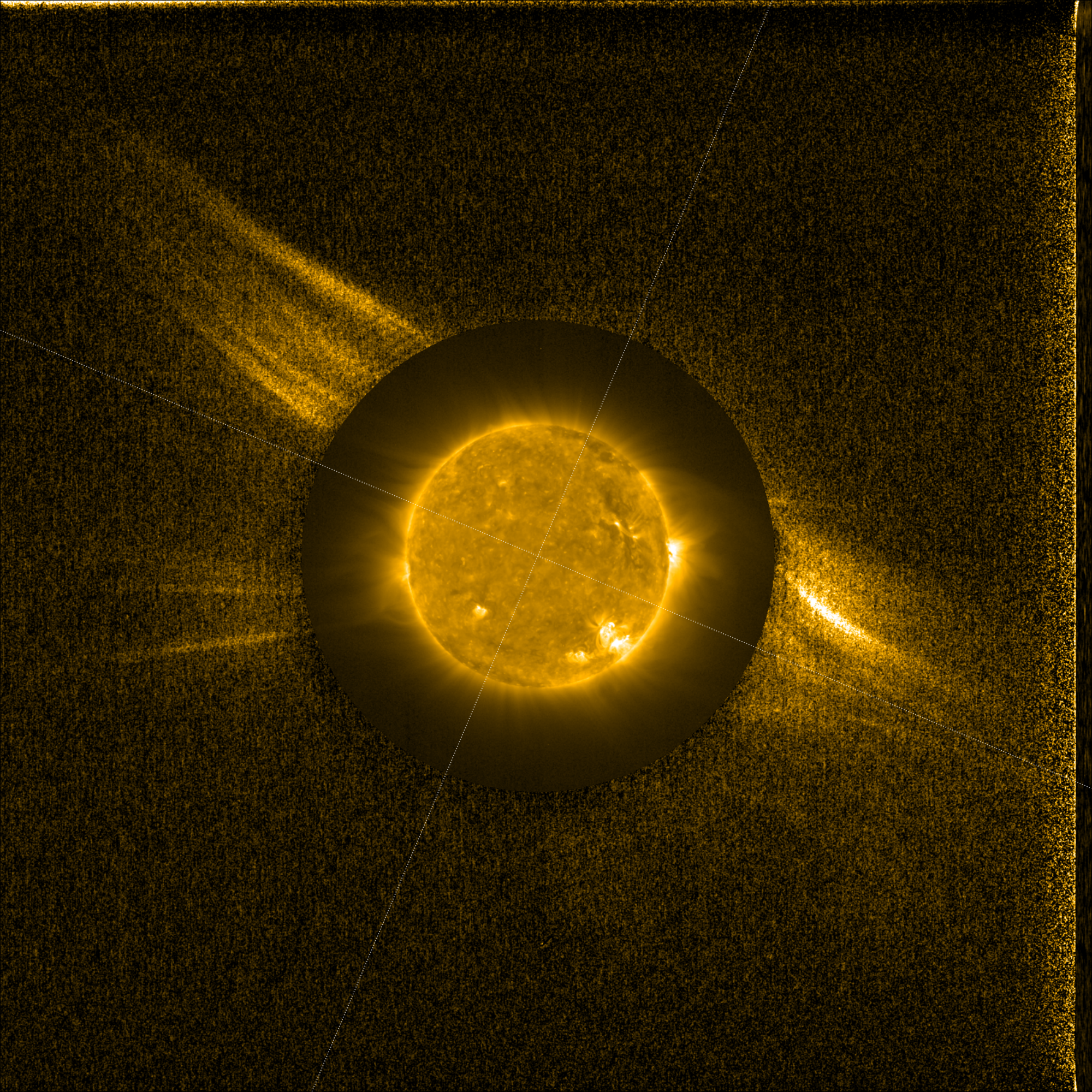}
    \caption{Composite of FSI \SI{17.4}{\nano\metre} images taken in disk mode (below 1.81~\rsun, 2021 September 8 at 23:55\,UT) and coronagraphic mode (September 9 at 00:42:03\,UT). As in all subsequent figures, the axes of the helio-projective coordinate system are plotted to materialize the roll angle of the spacecraft.}
    \label{fig:fsi_20210908}
\end{figure}

\begin{figure}
    \centering
    \includegraphics[width=\columnwidth]{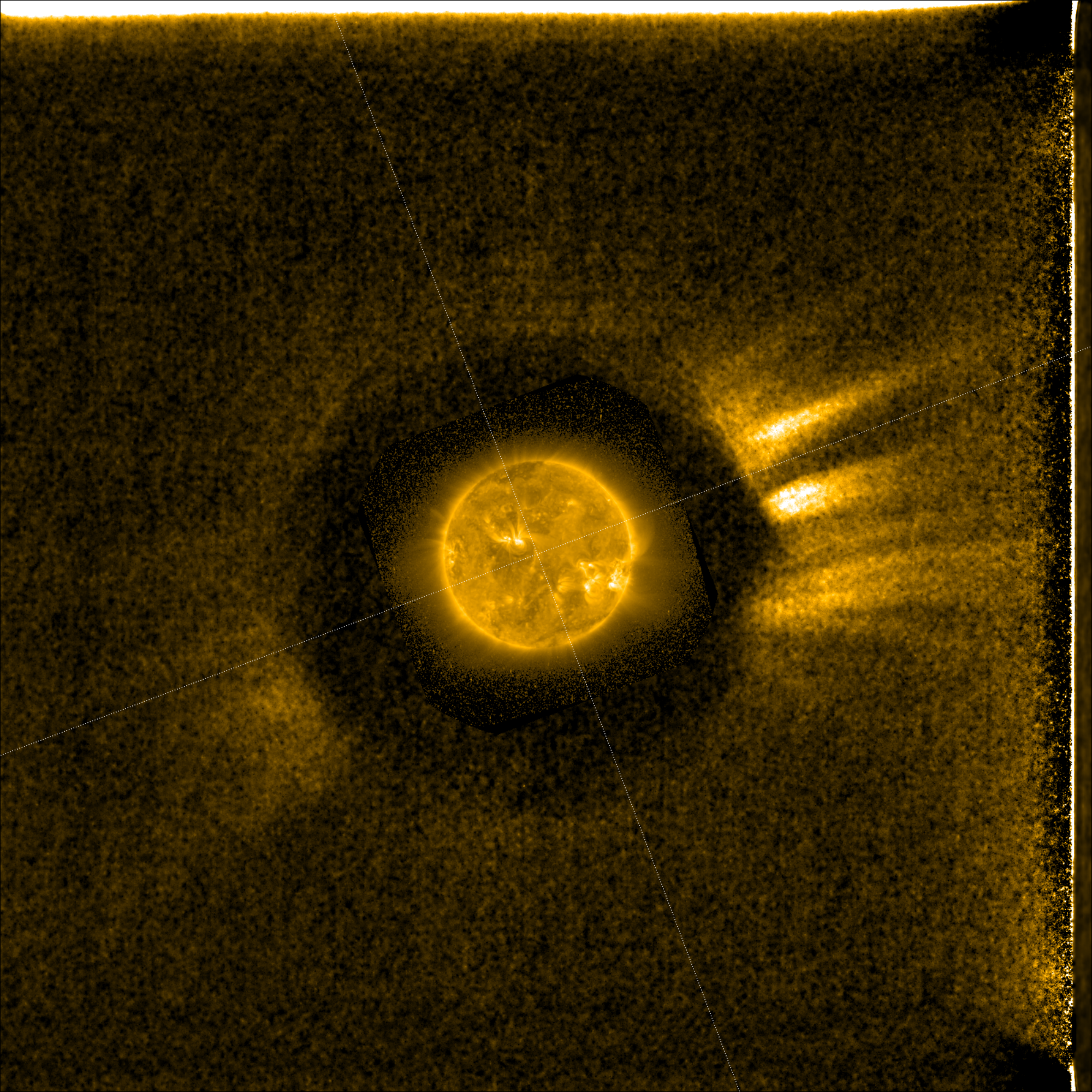}
    \caption{Composite of a SWAP \SI{17.4}{\nano\metre} image (below 2~\rsun, 2021 November 1 at 10:43\,UT) with an FSI \SI{17.4}{\nano\metre} image in coronagraphic mode (10:42\,UT).}
    \label{fig:fsi_202111}
\end{figure}

\begin{figure*}
    \centering
    \includegraphics[width=\textwidth]{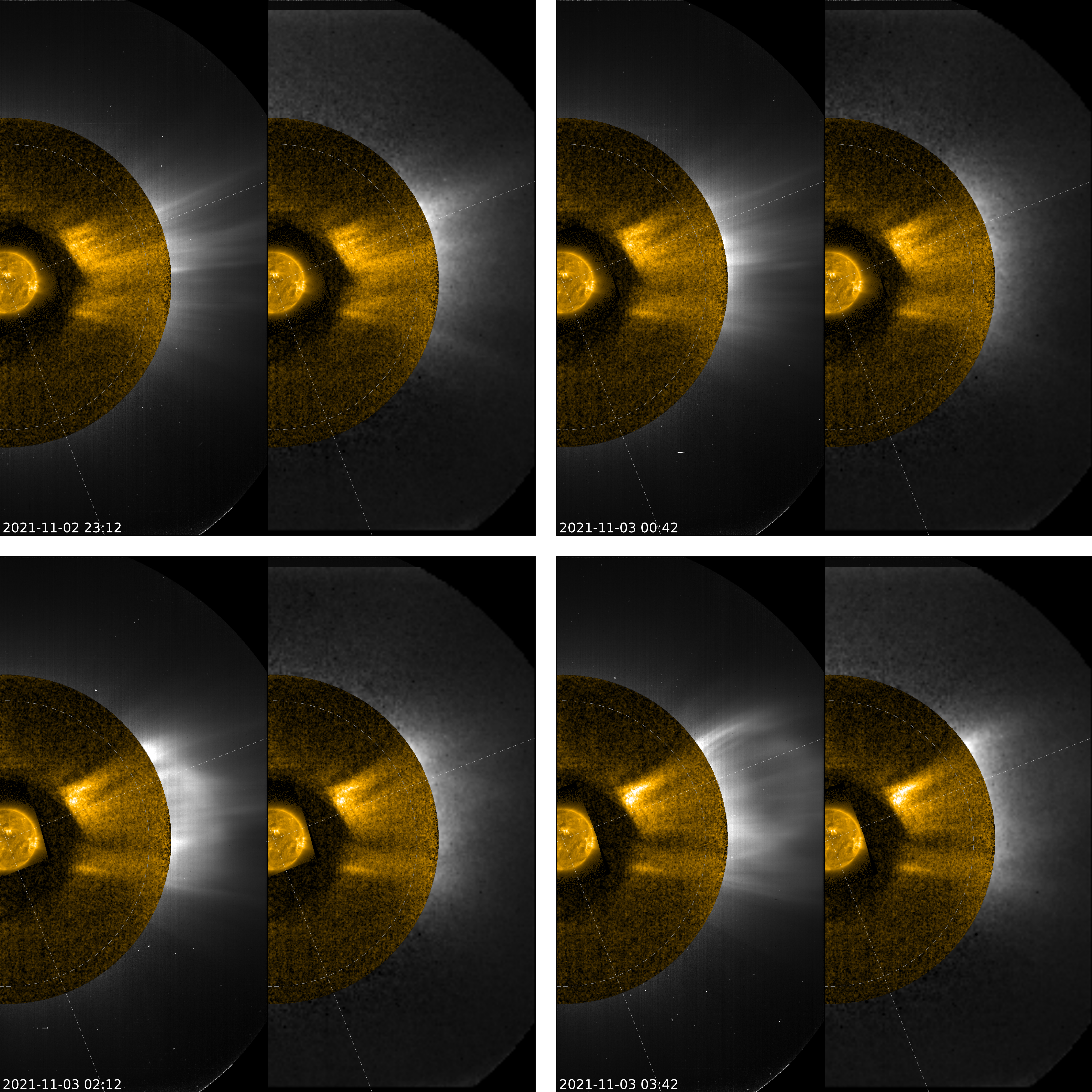}
    \caption{Propagation of a CME across the FSI and Metis FOVs. Left and right sides of each panel show a  comparison of SWAP (below 2~\rsun) and FSI \SI{17.4}{\nano\metre} (up to 5.6~\rsun) with Metis VL and \lya data, respectively. The dashed circle marks the position of the edge of the Metis occulting disk. An animated version of the figure is available online.}
    \label{fig:fsi_metis_202111}
\end{figure*}

\begin{figure}
    \centering
    \includegraphics[width=\columnwidth]{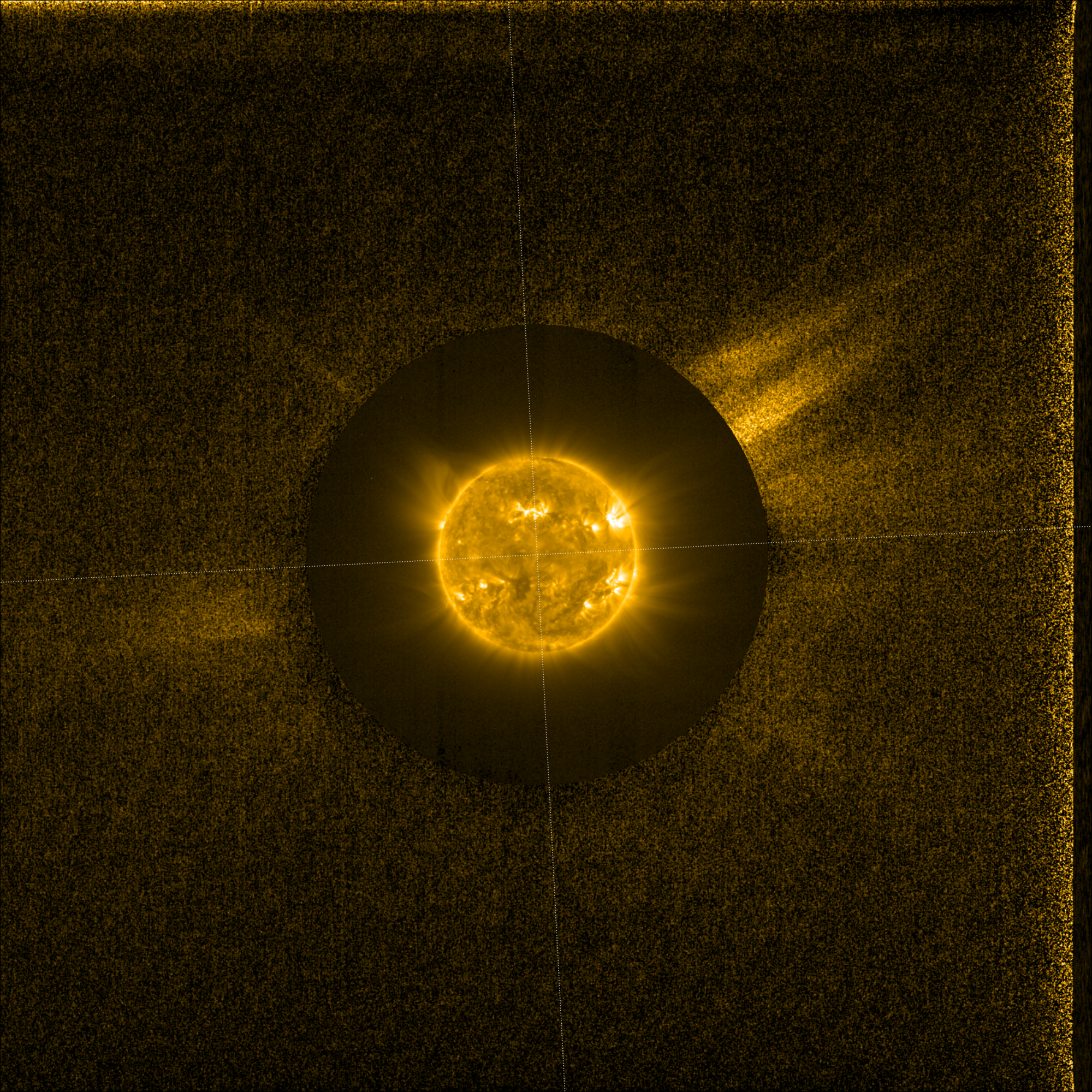}
    \caption{Composite of FSI \SI{17.4}{\nano\metre} images taken in disk mode (below 2.38~\rsun, 2022 February 8 at 12:15 UT) and coronagraphic mode (07:45 UT).}
    \label{fig:fsi_20220208}
\end{figure}

\subsection{September 2021 campaign\label{sec:campaign092021}}

This first campaign was run while the mission was still in cruise phase. A sequence of 50 images was obtained from 2021 September 9 at 00:42\,UT to 2021 September 9 at 09:30 with \SI{640}{\second} exposures at \SI{17.4}{\nano\metre}. Figure~\ref{fig:fsi_20210908} shows of one of the resulting images composited\footnote{Composite images in this paper were created using SunPy~\citep{sunpy_community2020} and AstroPy~\citep{astropy:2022}.} with an image taken 48 minutes before in disk mode. All EUV disk images in this paper are displayed with a $1/\gamma$ power scaling, with $\gamma=4$. All FSI coronagraph images are displayed with a linear scaling. Quasi-radial striations in the streamer above the north-east limb are similar to the fine structure of streamers reported by \citet{Ko2022}.

\subsection{November 2021 campaign\label{sec:campaign112021}}

Shortly after a major EUI onboard software update (2021 October 27), a longer campaign was attempted, while the mission was still in cruise phase. 
Altogether, 129 images were acquired at \SI{17.4}{\nano\metre} from 2021 November 1 at 00:42\,UT to 2021 November 3 at 23:42 and 38 images were acquired at \SI{30.4}{\nano\metre} on 2021 November 4 from 00:12\,UT to 21:12\,UT. Following an analysis of the data from the previous campaign, the exposure time was increased to \SI{1000}{\second}. The \SI{30.4}{\nano\metre} images were affected by the above-mentioned parasitic pattern and are not suited for scientific analysis. For an unknown reason, the subtraction of the matching exposure dark frame did not satisfactorily suppress the dark signal in the \SI{17.4}{\nano\metre} images for this period. In addition to the processing steps described in Sect.~\ref{sec:campaign112021},  from each frame we subtracted a background image corresponding (at each pixel) to the third percentile of intensity over the sequence. Figure~\ref{fig:fsi_202111} shows a composite between the first image of the campaign and a near simultaneous image from SWAP (Solar Orbiter was only \SI{2}{\degree} off the Sun-Earth line). At \SI{0.83}{\astronomicalunit} from the Sun the vignetting cutoff starts at 2.5~\rsun and there is no overlap with the SWAP FOV.

Figure~\ref{fig:fsi_metis_202111} shows composites with Metis VL (left half of each panel) and UV (right half) data at four different dates during the propagation of a coronal mass ejection (CME) over the west limb. Metis has observed other CMEs simultaneously in VL and at \lya~\citep{Andretta2021, Bemporad2022}, but it is the first time that simultaneous imaging up to 5.6~\rsun at \SI{17.4}{\nano\metre} is available. The expansion of the front is clearly seen in the first two frames in FSI before it reaches the Metis FOV. For a typical CME velocity of \SI{400}{\km\per\second}, the motion blur is 0.57 and 0.96~\rsun in the plane of the sky during the  1000 and \SI{1680}{\second} exposure times used by FSI and Metis, respectively. This is only \SIrange{10}{17}{\percent} of the CME's height in the top right panel, which explains why it is visible despite the very long exposure time. The match of the substructures of the CME is quite good between the EUV and VL, as can be seen in two bottom panels. The Metis UV data exhibits less structuring than VL, possibly because it is noisier, which makes the comparison more difficult.

\begin{figure*}
    \setlength\tabcolsep{2 pt}
    \centering
    \begin{tabular}{cc}
    \includegraphics[width=\columnwidth]{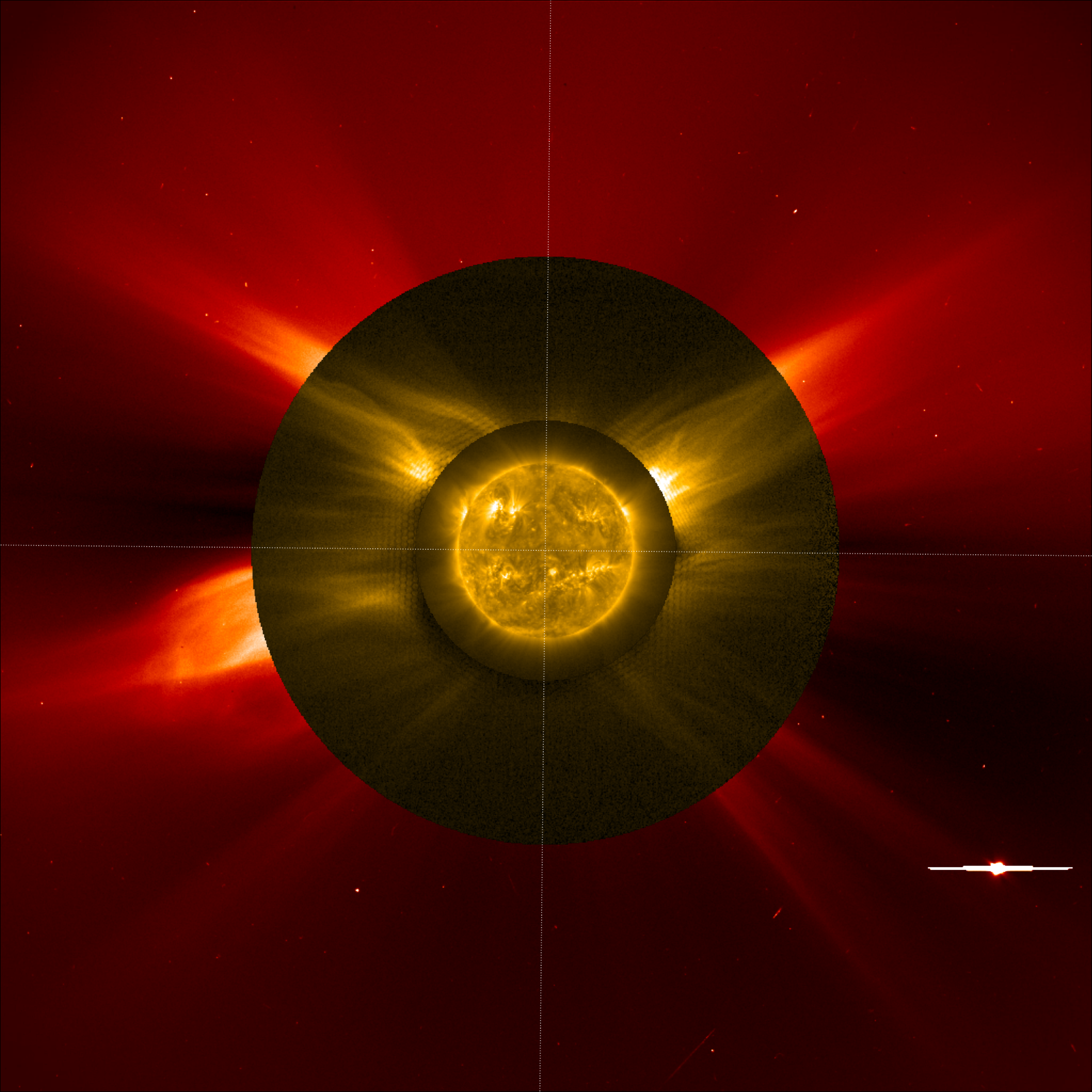} &
    \includegraphics[width=\columnwidth]{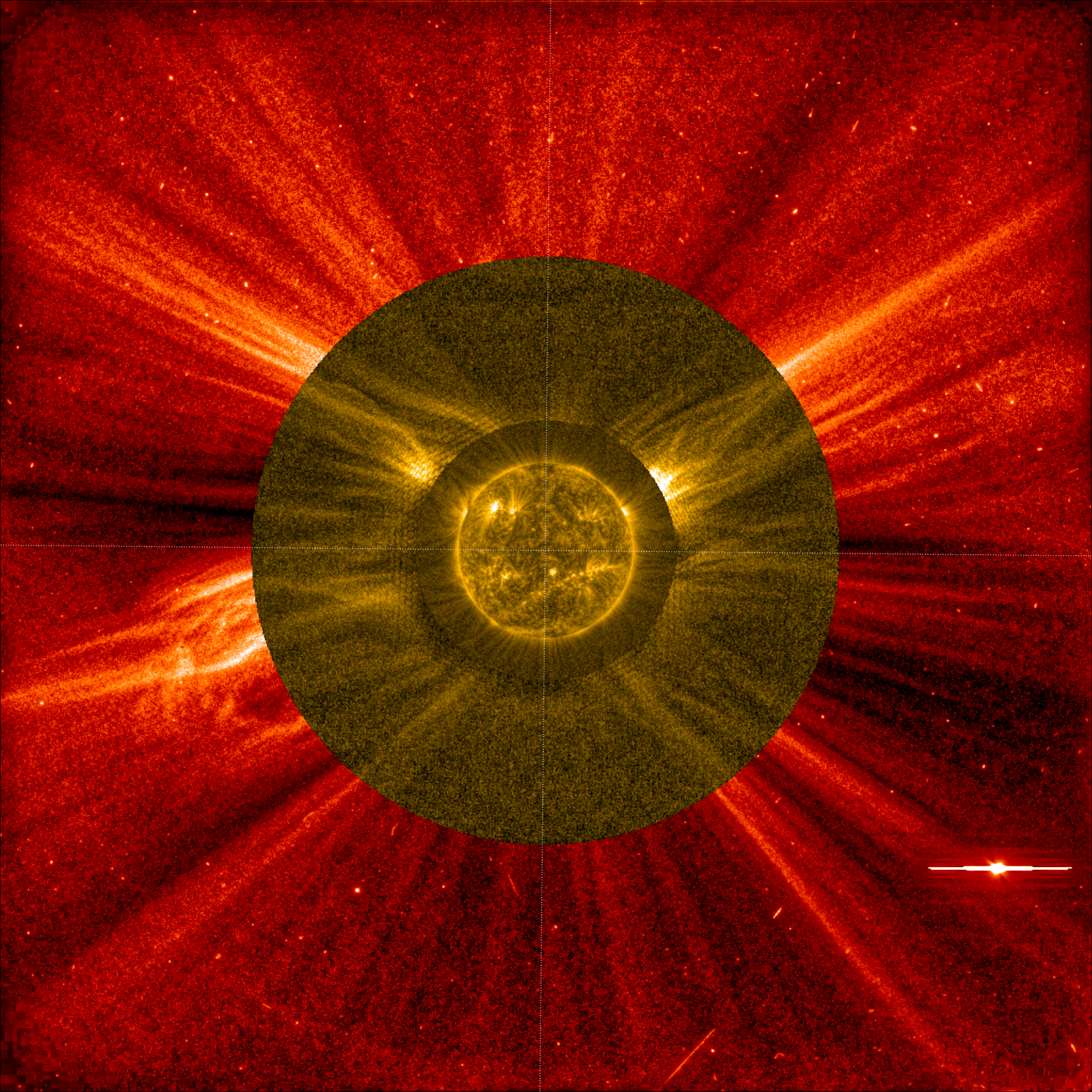}\\
    \end{tabular}
    \caption{Composites of images at \SI{17.4}{\nano\metre} from FSI in disk mode (below 1.5~\rsun, 11:29\,UT) and coronagraph mode (below 3.4~\rsun, 2022 March 7 16:00\,UT) and in VL from \lasco (16:12\,UT). Left: Original \lasco and FSI images, displayed using linear and square root scaling, respectively. Right: Images enhanced with the WOW filter.}
    \label{fig:fsi_lasco_20220307}
\end{figure*}

\begin{figure}
    \centering
    \begin{tabular}{c}
    \includegraphics[width=\columnwidth]{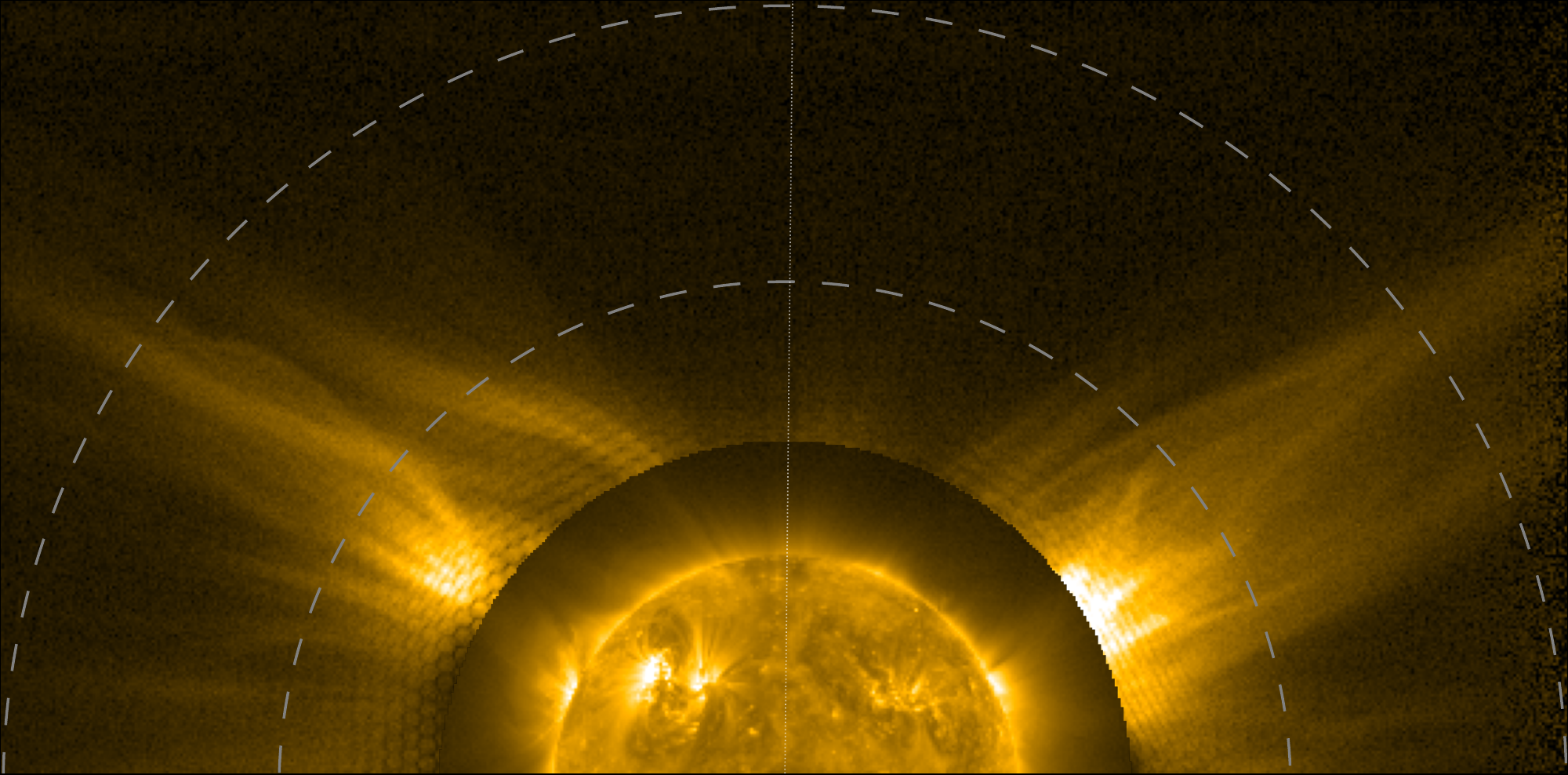}\\
    \includegraphics[width=\columnwidth]{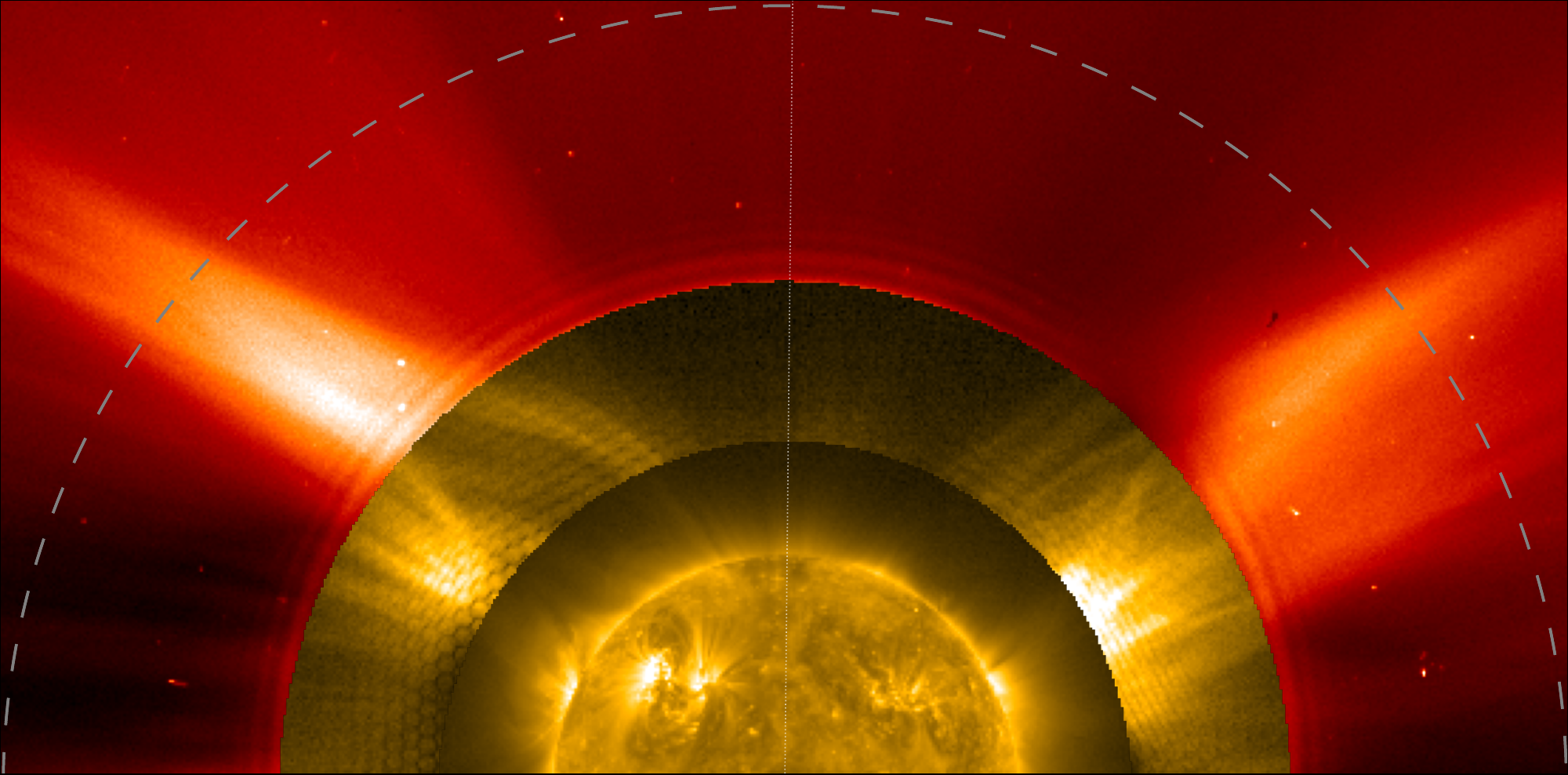}\\
    \end{tabular}
    \caption{Comparison of the northern half of the overlapping region between FSI and \lasco on 2022 March 7. Top: FSI composite (same images as in Fig.~\ref{fig:fsi_lasco_20220307}), extending to the edges of the detector. The outer dashed circle corresponds to the 3.4~\rsun boundary between FSI and \lasco used in Fig.~\ref{fig:fsi_lasco_20220307}. The inner dashed circle corresponds to the boundary between FSI and \lasco used in the bottom panel. Bottom: Same as above, with \lasco above 2.2~\rsun, the inner limit of the useful \lasco FOV.}
    \label{fig:fsi_lasco_20220307_detail}
    \end{figure}

\subsection{February 2022 campaign\label{sec:campaign022022}}

This short campaign on 2022 February 8 from 04:15\,UT to 07:45\,UT was a test run (and thus technically identical) to the subsequent 2022 March campaign discussed below.  Only 8 \SI{17.4}{\nano\metre} exposures were acquired of \SI{1000}{\second} each taken every 30 minutes. Solar Orbiter was \SI{0.79}{\astronomicalunit} from the Sun. Figure~\ref{fig:fsi_20220208} shows the last image of the sequence composited with a disk image taken 4 hours and 30 minutes later. The quasi-linear ray visible over the north-east limb in the outer image corresponds to the tip of a helmet streamer in the disk image.

\subsection{March 2022 campaign\label{sec:campaign032022}}

This campaign was run on 2022 March 7 from 16:00\,UT to 19:30\,UT with \SI{17.4}{\nano\metre} exposures of \SI{1000}{\second} each taken every 30 minutes, in support of the second flight of the Herschel sounding rocket, while Solar Orbiter was \SI{0.5}{\astronomicalunit} from the Sun. The FSI images were intended to provide the constraints on the coronal temperature required for the derivation of the abundance of Helium from the measurements made by the Herschel payload. The rocket did not return useful data, but since this campaign was run closer to the Sun than the others, FSI imaged brighter regions of the corona and revealed interesting structures up to the edge of the FOV (see Fig.~\ref{fig:fsi_lasco_20220307} and top panel of Fig.~\ref{fig:fsi_lasco_20220307_detail}). The morphology of the streamer above the north-west limb resembles the classical bipolar streamer magnetic field model by \citet{Pneuman1971}. The beginning of the sequence caught the trailing edge of a CME, clearly visible in the south-east in \lasco (Fig.~\ref{fig:fsi_lasco_20220307}).

There is overall a good match between the structures visible in \lasco and in FSI, both in the original images (left panel of Fig.~\ref{fig:fsi_lasco_20220307}) and those enhanced with the wavelets-optimized whitening filter~\citep[WOW, right panel of Fig.~\ref{fig:fsi_lasco_20220307},][]{Auchere2023}. There are some significant differences, however. Radial polar plumes revealed in the WOW-enhanced \lasco image do not have a counterpart in FSI beyond 1.4~\rsun. Plumes are fully visible in \ion{Fe}{X} \SI{789.2}{\nano\metre} and \ion{Fe}{XI} \SI{789.2}{\nano\metre} out to 2~\rsun during eclipses\citep{Habbal2011}. It is possible that they are less visible in the FSI images because of the lower signal-to-noise ratio. The FOV of FSI overlaps with that of \lasco above 2.2~\rsun (inner dashed circle in the top panel of Fig.~\ref{fig:fsi_lasco_20220307_detail})\footnote{The dark hexagonal pattern visible near the FSI vignetting cutoff is caused by the mesh grids supporting the front and focal filters (Fig.~\ref{fig:fsi_layout}). As described in~\citet{Auchere2011}, the modulation pattern cancels out in the nominal configuration because the footprint of the beam on the filters is an integer multiple of one grid tile. This condition is not satisfied with the occulting disk in place in the inner part of the FOV where the beam is partially vignetted.}. In this region, the north-east and north-west streamers exhibit different morphologies in VL and in the EUV. Morphological differences between emission lines and broadband VL have already been described by, for example, \cite{Habbal2007, Mierla2007}. There are several possible causes for the observed differences. Thomson scattering, which is responsible for the emission observed in VL, shows the free electrons and is proportional to their number density, while the emission lines dominating the FSI passband are emitted by multiply ionized iron atoms following collisional excitation (although resonant scattering may also play a role), which is proportional to the square of the electron number density. Density variations in the corona therefore produce more contrasted features in EUV images than in VL. For the same reason, the line of sight integration path is shorter in the EUV, resulting in fewer structures being superimposed. The intensity of emission lines is also a sharply peaked function of the temperature. Temperature variations may be responsible for, for example, the dark lane visible in the north-eastern streamer in the EUV but not in VL. There was a \SI{3}{\degree} separation between the lines of sight of the two instruments, which could explain small differences. A monthly average was subtracted from the \lasco data and this could possibly produce gradients that are not present in the EUV images. Motion blur (mostly radial) can also affect the FSI images due to the \SI{1000}{\second} exposure time used.

\subsection{December 2022 campaign\label{sec:campaign122022}}

\begin{figure}
    \centering
    \includegraphics[width=\columnwidth]{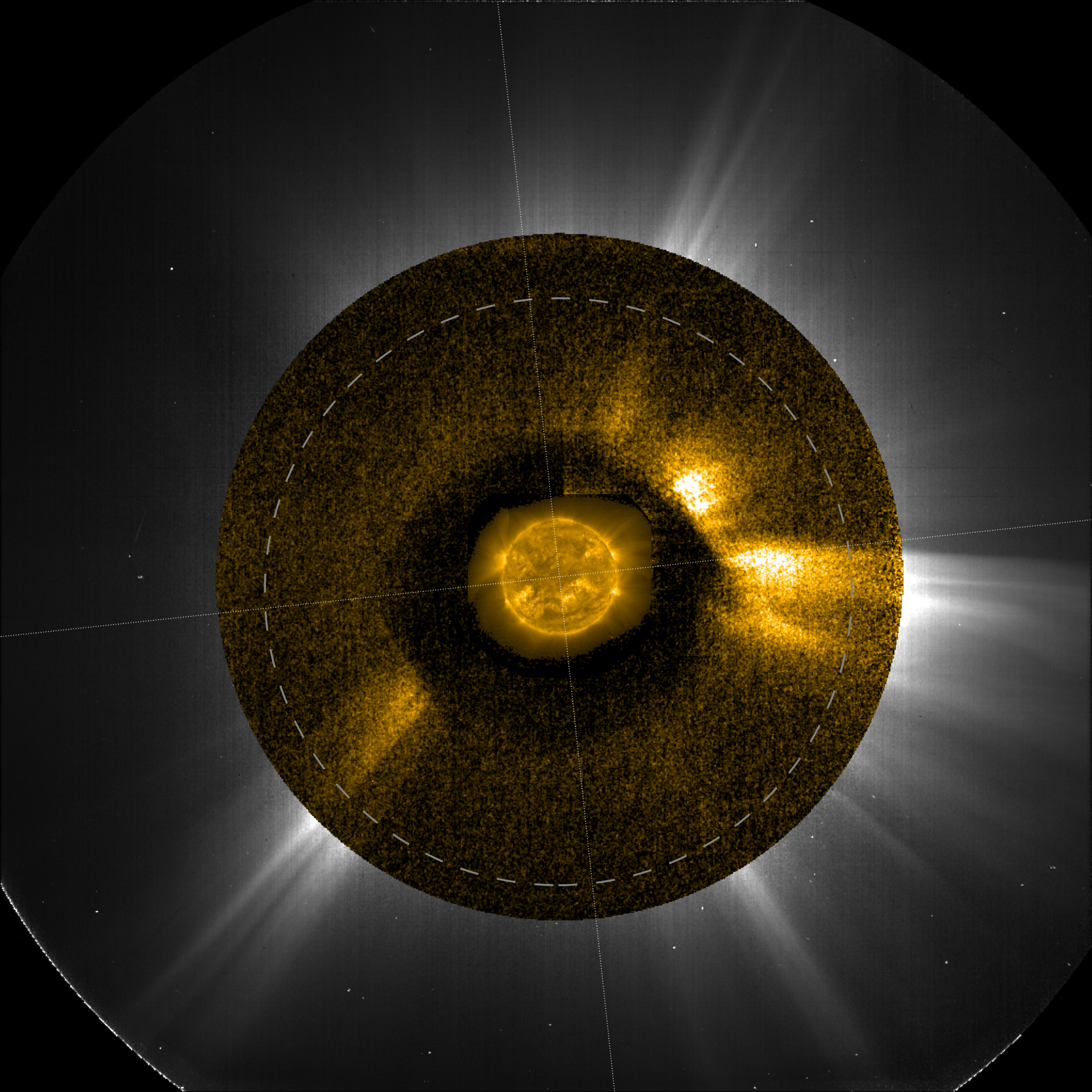}
    \caption{Composite of images taken on 2022 December 15: EUVI-A (below 1.9~\rsun, 18:34\,UT), FSI \SI{17.4}{\nano\metre} (below 6.1~\rsun, 18:30\,UT) and Metis VL (18:30\,UT). The dashed circle marks the position of the edge of the Metis occulting disk. An animated version of this figure is available online.}
    \label{fig:fsi_202212}
\end{figure}

The campaign started on 2022 December 5 at 04:00\,UT and ended on 2023 January 1 at 21:36\,UT. It consisted of \SI{17.4}{\nano\metre} exposures of \SI{1000}{\second,} each taken every 30 minutes. It was interrupted on December 11 from 03:00 to 08:00\,UT to support quadrature observations with Parker Solar Probe. It is the longest FSI coronagraph mode campaign run to date and includes 1547 images. Figure~\ref{fig:fsi_202212} shows a composite image of FSI and Metis VL data, with a disk image from EUVI-A, which was the closest EUV imager (Table~\ref{tab:campaigns}). As in Fig.~\ref{fig:fsi_lasco_20220307}, some of the radial structures in the Metis FOV do not have a counterpart in FSI, either because they are below the detection limit or because of the different formation processes among the observed emissions. In the animated version of this figure covering the whole sequence, several CMEs can be tracked through the FSI and Metis FOVs. The bright linear rays visible in the animation line up with the edge of streamers. They likely appear thinner than in VL because the contrast in the EUV is increased due to the proportionality of the intensity to the square of the electron number density.

\section{Conclusions\label{sec:conclusions}}

We present observations carried out thus far by FSI in coronagraphic mode at \SI{17.4}{\nano\metre}. Going up to 6.4~\rsun at the detector's edge, these are the widest field EUV images of the corona made to date. Without the presence of the occulting disk, stray light amounts to 90\% of the signal at 2~\rsun to 99\% at 5~\rsun. The stray light-free images provided by FSI in the coronagraph mode enable photometric studies of the distribution of intensity of the observed emission lines. Some items of particular interest will be the comparison with eclipse observations of other emission lines ~\citep{Boe2022}, the quantification of the contribution of resonant scattering as a function of height, and the joint inversion of Metis and FSI data to constrain the coronal temperature in the regions of overlap~\citep{Abbo2023}. More coronagraph mode campaigns are planned in the near future, with a focus on periods during which the separation angle with other EUV disk imagers is small. The instrument will also provide novel images of the corona later in the mission from out of ecliptic viewpoints.

While this paper demonstrates the possibility of wide field coronagraphic EUV imaging, FSI was not optimized for this mode of observation. The entrance pupil was sized to ensure the image quality and to limit the heat flux entering the instrument while allowing sufficient signal on disk in \SI{10}{\second} exposures. Furthermore, the occulting disk was installed late in the project life at a sub-optimal position. With minor modifications, the efficiency of an FSI-based coronagraph could be increased by two orders of magnitude, which would allow images similar to those presented here to be acquired in \SI{10}{\second}. This opens up the perspective of a single EUV telescope able to monitor the coronal activity from Sun center to 6~\rsun. Compared to a VL coronagraph, an EUV instrument offers several advantages. There is no background emission from scattering off dust (F-corona). The contrast between the disk and the corona is several orders of magnitude smaller in the EUV than in VL, and since the corresponding wavelengths are around 20 times shorter, diffraction by the edges is reduced. Instrumental stray light is thus easier to control. Indeed, a single circular occulting disk is sufficient in FSI, while complex multiple stage systems are necessary in VL -- and without the ability to completely suppress stray light. This also makes an EUV coronagraph less demanding in terms of platform pointing accuracy and stability. The shorter line of sight in EUV images compared to VL may possibly make the EUV coronagraph less sensitive to halo CMEs -- which are of particular interest for space weather applications; however, the possibility that resonant scattering plays a significant role at large distances in spectral lines may allow for their detection. 

While EUV coronagraphs offer an alternative solution, the two types of instruments are in fact complementary, with VL providing the electron number density and the EUV offering access to the emission measure as well as the plasma temperature, provided that several passbands are available. This would provide unprecedented diagnostic capability for the coronal plasma beyond 2~\rsun.

\begin{acknowledgements}
The author would like to thank Pierre Rochus, Principal Investigator of EUI until the launch, for letting him add the occulting disk to the door design as an undocumented feature in 2014, after CDR.

Solar Orbiter is a space mission of international collaboration between ESA and NASA, operated by ESA. The EUI instrument was built by CSL, IAS, MPS, MSSL/UCL, PMOD/WRC, ROB, LCF/IO with funding from the Belgian Federal Science Policy Office (BELSPO/PRODEX PEA C4000134088); the Centre National d’Etudes Spatiales (CNES); the UK Space Agency (UKSA); the Bundesministerium für Wirtschaft und Energie (BMWi) through the Deutsches Zentrum für Luft- und Raumfahrt (DLR); and the Swiss Space Office (SSO). The ROB team thanks the Belgian Federal Science Policy Office (BELSPO) for the provision of financial support in the framework of the PRODEX Programme of the European Space Agency (ESA) under contract numbers 4000134474 and 4000136424.

Solar Orbiter is a space mission of international collaboration between ESA and NASA, operated by ESA. Metis was built and operated with funding from the Italian Space Agency (ASI), under contracts to the National Institute of Astrophysics (INAF) and industrial partners. Metis was built with hardware contributions from Germany (Bundesministerium für Wirtschaft und Energie through DLR), from the Czech Republic (PRODEX) and from ESA.

SOHO/LASCO data are produced by a consortium of the Naval Research Laboratory (USA), Max-Planck-Institut für Aeronomie (Germany), Laboratoire d'Astronomie (France), and the University of Birmingham (UK). SOHO is a project of international cooperation between ESA and NASA. 
\end{acknowledgements}

\bibliographystyle{aa}

\bibliography{bibliography}

\end{document}